\documentstyle[twoside]{article}
\catcode`\@=11
\long\def\@makefntext#1{
\protect\noindent \hbox to 3.2pt {\hskip-.9pt
$^{{\eightrm\@thefnmark}}$\hfil}#1\hfill}               

\def\thefootnote{\fnsymbol{footnote}}
\def\@makefnmark{\hbox to 0pt{$^{\@thefnmark}$\hss}}    

\def\ps@myheadings{\let\@mkboth\@gobbletwo
\def\@oddhead{\hbox{}
\rightmark\hfil\eightrm\thepage}
\def\@oddfoot{}\def\@evenhead{\eightrm\thepage\hfil
\leftmark\hbox{}}\def\@evenfoot{}
\def\sectionmark##1{}\def\subsectionmark##1{}}

\oddsidemargin=\evensidemargin
\renewcommand{\thefootnote}{\fnsymbol{footnote}}

\newcounter{sectionc}\newcounter{subsectionc}\newcounter{subsubsectionc}
\renewcommand{\section}[1] {\vspace{12pt}\addtocounter{sectionc}{1}
\setcounter{subsectionc}{0}\setcounter{subsubsectionc}{0}\noindent
        {\tenbf\thesectionc. #1}\par\vspace{5pt}}
\renewcommand{\subsection}[1] {\vspace{12pt}\addtocounter{subsectionc}{1}
        \setcounter{subsubsectionc}{0}\noindent
        {\bf\thesectionc.\thesubsectionc. {\kern1pt \bfit #1}}\par\vspace{5pt}}
\renewcommand{\subsubsection}[1] {\vspace{12pt}\addtocounter{subsubsectionc}{1}
        \noindent{\tenrm\thesectionc.\thesubsectionc.\thesubsubsectionc.
        {\kern1pt \tenit #1}}\par\vspace{5pt}}
\newcommand{\nonumsection}[1] {\vspace{12pt}\noindent{\tenbf #1}
        \par\vspace{5pt}}

\newcounter{appendixc}
\newcounter{subappendixc}[appendixc]
\newcounter{subsubappendixc}[subappendixc]
\renewcommand{\thesubappendixc}{\Alph{appendixc}.\arabic{subappendixc}}
\renewcommand{\thesubsubappendixc}
        {\Alph{appendixc}.\arabic{subappendixc}.\arabic{subsubappendixc}}

\renewcommand{\appendix}[1] {\vspace{12pt}
        \refstepcounter{appendixc}
        \setcounter{figure}{0}
        \setcounter{table}{0}
        \setcounter{lemma}{0}
        \setcounter{theorem}{0}
        \setcounter{corollary}{0}
        \setcounter{definition}{0}
        \setcounter{equation}{0}
        \renewcommand{\thefigure}{\Alph{appendixc}.\arabic{figure}}
        \renewcommand{\thetable}{\Alph{appendixc}.\arabic{table}}
        \renewcommand{\theappendixc}{\Alph{appendixc}}
        \renewcommand{\thelemma}{\Alph{appendixc}.\arabic{lemma}}
        \renewcommand{\thetheorem}{\Alph{appendixc}.\arabic{theorem}}
        \renewcommand{\thedefinition}{\Alph{appendixc}.\arabic{definition}}
        \renewcommand{\thecorollary}{\Alph{appendixc}.\arabic{corollary}}
        \renewcommand{\theequation}{\Alph{appendixc}.\arabic{equation}}
        \noindent{\tenbf Appendix \theappendixc #1}\par\vspace{5pt}}
\newcommand{\subappendix}[1] {\vspace{12pt}
        \refstepcounter{subappendixc}
        \noindent{\bf Appendix \thesubappendixc. {\kern1pt \bfit #1}}
        \par\vspace{5pt}}
\newcommand{\subsubappendix}[1] {\vspace{12pt}
        \refstepcounter{subsubappendixc}
        \noindent{\rm Appendix \thesubsubappendixc. {\kern1pt \tenit #1}}
        \par\vspace{5pt}}

\newcommand{\textlineskip}{\baselineskip=13pt}
\newcommand{\smalllineskip}{\baselineskip=10pt}

\def\eightcirc{
\begin{picture}(0,0)
\put(4.4,1.8){\circle{6.5}}
\end{picture}}
\def\eightcopyright{\eightcirc\kern2.7pt\hbox{\eightrm c}}

\newcommand{\copyrightheading}[1]
        {\vspace*{-2.5cm}\smalllineskip{\flushleft
        {\footnotesize $\eightcopyright$\, World Scientific Publishing
         Company}\\
         }}

\def\abstracts#1#2#3{{
        \centering{\begin{minipage}{4.5in}\baselineskip=10pt\footnotesize
        \parindent=0pt #1\par
        \parindent=15pt #2\par
        \parindent=15pt #3
        \end{minipage}}\par}}

\newcommand{\bibit}{\nineit}
\newcommand{\bibbf}{\ninebf}
\renewenvironment{thebibliography}[1]
        {\frenchspacing
         \ninerm\baselineskip=11pt
         \begin{list}{\arabic{enumi}.}
        {\usecounter{enumi}\setlength{\parsep}{0pt}
         \setlength{\leftmargin 12.7pt}{\rightmargin 0pt} 
         \setlength{\itemsep}{0pt} \settowidth
        {\labelwidth}{#1.}\sloppy}}{\end{list}}

\newcounter{itemlistc}
\newcounter{romanlistc}
\newcounter{alphlistc}
\newcounter{arabiclistc}

\newcommand{\fcaption}[1]{
        \refstepcounter{figure}
        \setbox\@tempboxa = \hbox{\footnotesize Fig.~\thefigure. #1}
        \ifdim \wd\@tempboxa > 5in
           {\begin{center}
        \parbox{5in}{\footnotesize\smalllineskip Fig.~\thefigure. #1}
            \end{center}}
        \else
             {\begin{center}
             {\footnotesize Fig.~\thefigure. #1}
              \end{center}}
        \fi}

\newcommand{\tcaption}[1]{
        \refstepcounter{table}
        \setbox\@tempboxa = \hbox{\footnotesize Table~\thetable. #1}
        \ifdim \wd\@tempboxa > 5in
           {\begin{center}
        \parbox{5in}{\footnotesize\smalllineskip Table~\thetable. #1}
            \end{center}}
        \else
             {\begin{center}
             {\footnotesize Table~\thetable. #1}
              \end{center}}
        \fi}

\def\@citex[#1]#2{\if@filesw\immediate\write\@auxout
        {\string\citation{#2}}\fi
\def\@citea{}\@cite{\@for\@citeb:=#2\do
        {\@citea\def\@citea{,}\@ifundefined
        {b@\@citeb}{{\bf ?}\@warning
        {Citation `\@citeb' on page \thepage \space undefined}}
        {\csname b@\@citeb\endcsname}}}{#1}}

\newif\if@cghi
\def\cite{\@cghitrue\@ifnextchar [{\@tempswatrue
        \@citex}{\@tempswafalse\@citex[]}}
\def\citelow{\@cghifalse\@ifnextchar [{\@tempswatrue
        \@citex}{\@tempswafalse\@citex[]}}
\def\@cite#1#2{{$\null^{#1}$\if@tempswa\typeout
        {IJCGA warning: optional citation argument
        ignored: `#2'} \fi}}

\def\pmb#1{\setbox0=\hbox{#1}
        \kern-.025em\copy0\kern-\wd0
        \kern.05em\copy0\kern-\wd0
        \kern-.025em\raise.0433em\box0}

\def\fnt#1#2{\footnotetext{\kern-.3em
        {$^{\mbox{\scriptsize #1}}$}{#2}}}

\def\runninghead#1#2{\pagestyle{myheadings}
\markboth{{\protect\footnotesize\it{\quad #1}}\hfill}
{\hfill{\protect\footnotesize\it{#2\quad}}}}
\font\tenrm=cmr10
\font\tenit=cmti10
\font\tenbf=cmbx10
\font\bfit=cmbxti10 at 10pt
\font\ninerm=cmr9
\font\nineit=cmti9
\font\ninebf=cmbx9
\font\eightrm=cmr8

\def\qed{\hbox{${\vcenter{\vbox{                        
   \hrule height 0.4pt\hbox{\vrule width 0.4pt height 6pt
   \kern5pt\vrule width 0.4pt}\hrule height 0.4pt}}}$}}

\renewcommand{\thefootnote}{\fnsymbol{footnote}}        

\def\bsc{{\sc a\kern-6.4pt\sc a\kern-6.4pt\sc a}}       
\def\bflatex{\bf L\kern-.30em\raise.3ex\hbox{\bsc}\kern-.14em
T\kern-.1667em\lower.7ex\hbox{E}\kern-.125em X}

\def\del{\partial}
\def\Vka{V_{\kappa}}
\def\Vla{V_{\lambda}}
\def\Vmu{V_{\mu}}

\def\Vkala{V_{\kappa\lambda}}
\def\Vkamu{V_{\kappa\mu}}

\def\Vlamu{V_{\lambda\mu}}

\def\Vkalamu{V_{\kappa\lambda\mu}}

\def\Vkalamunu{V_{\kappa\lambda\mu\nu}}
\begin{document}

\runninghead{Inverse Mass Expansions from Worldline Path Integrals}{
Inverse Mass Expansions from Worldline Path Integrals}

\normalsize\textlineskip
\thispagestyle{empty}
\setcounter{page}{1}


\vspace*{-36pt}

\hfill HD-THEP-95-20

\hfill hep-th/9505077

\centerline{\bf }
\centerline{\bf INVERSE MASS EXPANSIONS FROM WORLDLINE PATH INTEGRALS ---}
\vspace*{0.035truein}
\centerline{\bf HIGHER ORDER COEFFICIENTS AND ORDERING PROBLEMS
\footnote{Talk given by D.~Fliegner at the Fourth International Workshop on
Software Engineering and Artificial Intelligence for High Energy and Nuclear
Physics AIHENP95, Pisa (Italy), Apr 3-8 1995}
}

\vspace*{0.37truein}
\centerline{\footnotesize D.~FLIEGNER, P.~HABERL, M.G.~SCHMIDT
}
\vspace*{0.015truein}
\centerline{\footnotesize\it Institut f\"ur Theoretische Physik,
Universit\"at Heidelberg}
\baselineskip=10pt
\centerline{\footnotesize\it Philosophenweg 16, D-69120 Heidelberg, Germany
}
\vspace*{10pt}
\centerline{\normalsize and}
\vspace*{10pt}
\centerline{\footnotesize C.~SCHUBERT}
\vspace*{0.015truein}
\centerline{\footnotesize\it Institut f\"ur Physik,
Humboldt Universit\"at zu Berlin}
\baselineskip=10pt
\centerline{\footnotesize\it Invalidenstr. 110, D-10115 Berlin, Germany}
\vspace*{0.225truein}
%
\abstracts{
Higher order coefficients of the inverse mass expansion of one--loop
effective actions are obtained from a one--dimensional path integral
representation. For the evaluation of the path integral with
Wick contractions a suitable
Green function has to be chosen. We consider the case of a
massive scalar loop in the background of both a scalar potential and
a (non--abelian) gauge field. For the pure scalar case the
method yields the coefficients of the expansion in a minimal set of
basis terms whereas complicated ordering problems arise in gauge theory.
An appropriate reduction scheme is discussed.}{}{}
\textheight=7.8truein
\setcounter{footnote}{0}
\renewcommand{\thefootnote}{\alph{footnote}}

\textlineskip                  
\vspace*{12pt}                 
\noindent
In recent years, the inverse mass expansion of one--loop effective
actions has been applied in fields as different as the calculation
of quark determinants, bubble nuc\-leation during the electroweak phase
transition or baryon number violation by instanton or sphaleron processes.
A variety of approaches and results can be found in the literature [1].
The actual computation of higher order coefficients, however, suffers from
practical limitations and was worked out only to order ${\cal O}(T^6)$ for
the ungauged theory [2] and ${\cal O}(T^5)$ for the gauged case [3].
In our approach, instead of using conventional heat--kernel techniques
we start from a representation of the one--loop effective action as a
one--dimensional path integral. For the evaluation of the path integral we
make use of recent progress in calculating one--loop amplitudes inspired by
string theory [4,5,6], which enables us to compute the inverse mass expansion
to high orders.
\par
Formally, the one-loop effective action for a scalar particle with mass
$m$ in the background of a gauge field $A_\mu$ (abelian or non--abelian)
and a matrix--valued scalar potential $V$ is obtained as the determinant
of a fluctuation operator $M$, which can be written in the Schwinger
proper time formalism as
\begin{equation}
\Gamma_{\rm eff}[A,V]=-{\rm ln}({\rm det M})=-{\rm Tr}({\rm ln} M)
=\int_0^\infty{{\rm d}T\over T}{\rm Tr}\;{\rm e}^{-T M}\;.
\label{Geff}
\end{equation}
For the case we are considering the operator $M$ reads
(in $d$ Euclidean dimensions)
\begin{equation}
M = - D^2 + m^2 + V(x) \qquad{\rm with}\qquad
D_\mu=\del_\mu - ig A_\mu (x)\,.
\end{equation}
The expansion of $\Gamma_{\rm eff}$ in inverse powers of the heavy
mass $m$ is obtained from expanding the exponential in Eq.\ (\ref{Geff})
in powers of the proper time $T$. (Since we will not evaluate the
proper time integration, we do not comment on its regularization.)
\par
The operator trace is nothing but the diagonal element of the heat
kernel for the operator $M$, integrated over space--time.
It can be represented as a one--dimensional
path integral on the space of closed loops in space-time
with fixed proper time circumference $T$.
The action then takes the form
\begin{equation}
\Gamma[A,V]=
\int_0^\infty {{\rm d}T\over T} {\rm e}^{-m^2T} {\rm tr} \;\;
{\cal P}\!\! \int_{x(T)=x(0)} \!\!\!\!\!\!\!\!\!\!\!\! {\cal D}x
\exp \Bigl[ - \int_0^T {\rm d}\tau \bigl( {\dot{x}^2\over 4}
+ ig \dot{x}^\mu A_\mu  +V(x) \bigr)\Bigr]\,,
\label{GAV}
\end{equation}
where ${\cal P}$ denotes path ordering and the operator trace ${\rm Tr}$
has reduced to an ordinary matrix trace ${\rm tr}$.
\par
For the Wick contractions one needs the Green function $G_B$ for the
Laplacian on the circle with periodic boundary conditions [5].
However, without the introduction of a ``background charge'' $\rho$ on
the wordline, the defining equation for $G_B$ has no solution.
A convenient choice is a uniformly distributed $\rho=1/T$,
which yields the modified defining equation
\begin{equation}
{1\over 2} {\del^2\over\del\tau_1^2} G_B(\tau_1,\tau_2)
= \delta(\tau_1-\tau_2) - \rho(\tau_1)
= \delta(\tau_1-\tau_2) - {1\over T}
\label{green}
\end{equation}
with the solution
\begin{equation}
G_B(\tau_1,\tau_2) = |\tau_1-\tau_2|-{(\tau_1-\tau_2)^2\over T}
\;.
\end{equation}
It is precisely this choice of the background charge which distinguishes
our approach from previous uses of path integrals in this context [7].
\par
As a consequence of the introduction of a background charge,
the Green function $G_B$ fulfills the identity
(integration by parts)
\[
\int_0^T {\rm d}\tau_2 \;{1\over 2} G_B(\tau_1,\tau_2)
\ddot{x}(\tau_2) = x(\tau_1) - {1\over T} \int_0^T
{\rm d}\tau_2 \;x(\tau_2)\,,
\]
where the second term on the r.h.s.\ should vanish.
This suggests that one should introduce a loop center of mass $x_0$
and a relative loop coordinate $y$,
\begin{equation}
x^{\mu}(\tau) = x^\mu_0 + y^\mu(\tau)
\qquad\mbox{with}\qquad
%
\int_0^T {\rm d}\tau \;y^{\mu}(\tau) = 0\,,
\end{equation}
and separate the integration over the center of mass from
the path integral:
\begin{equation}
\int {\cal D}x = \int {\rm d}^dx_0 \int {\cal D}y \;,
\end{equation}
resulting in a path integral over the space of all loops with a common
center of mass $x_0$. With $\dot{y}^\mu=\dot{x}^\mu$,
the inverse mass expansion of $\Gamma_{\rm eff}$ follows from
expanding the path ordered exponential in Eq.\ (\ref{GAV}).
\par
Choosing the background gauge field to be in Fock-Schwinger gauge
with reference point $x_0$ [6], it can be written as
\begin{equation}
A_\mu(x_0+y) = y^\rho \int_0^1 {\rm d}\eta \;\eta \;
F_{\rho\mu}(x_0+\eta y)\;.
\label{Acov}
\end{equation}
The main advantage of the Fock-Schwinger gauge is that the
Taylor expansion of the background fields around $x_0$
can be done in a covariant way,
\begin{eqnarray}
F_{\rho\mu}(x_0+\eta y) &=& {\rm e}^{\eta y D} F_{\rho\mu}(x_0) \;,
\nonumber\\
V(x_0+ y) &=& {\rm e}^{ y D} V(x_0) \;.
\label{Fcov}
\end{eqnarray}
With these ingredients, Eq.\ (3) takes the manifestly covariant
form
\begin{eqnarray}
\Gamma[F,V]\!\!&=&\!\!  \int_0^\infty {{\rm d}T\over T}
{\rm e}^{-m^2 T} \;{\rm tr}\int {\rm d}^d x_0 \sum_{n=0}^\infty
(-1)^n \int {\cal D}y \;{\rm exp} \Bigl[ -
\int_0^T \!\!\!{\rm d}\tau \;{\dot{y}^2\over 4} \Bigr]
\nonumber\\
&&\times\int_0^{T}\!{\rm d}\tau_1 \int_0^{\tau_1}
\!\!{\rm d}\tau_2 ... \int_0^{\tau_{n-1}} \!\!\!{\rm d}\tau_n
\prod_{j=1}^n \Bigl[\;{\rm e}^{ y(\tau_j)D_{(j)}} V^{(j)}(x_0)
\;+\nonumber\\
&&\qquad+ \; ig \dot{y}^{\mu_j}(\tau_j)
y^{\rho_j}(\tau_j)\int_0^1 {\rm d}\eta_j \eta_j \;
 {\rm e}^{\eta_j y(\tau_j) D_{(j)}}
F_{\rho_j\mu_j}^{(j)}(x_0)  \Bigr] \;.
\label{master}
\end{eqnarray}
\par
For the calculation of the coefficients of the inverse mass expansion to
a given order, the following steps have to be performed:
\par
{\it (i) Wick-contractions\/}: One truncates the sum in Eq.\ (\ref{master})
and expands the exponentials. All possible Wick contractions have to be
evaluated using the contraction rules
\begin{eqnarray}
\langle x^{\mu}(\tau_1) x^{\nu}(\tau_2) \rangle \!\!&=&\!\!
-g^{\mu\nu} G_B(\tau_1,\tau_2) =
-g^{\mu\nu} \Bigl[|\tau_1-\tau_2|-{(\tau_1-\tau_2)^2\over T}
\Bigr] \;,\nonumber\\
\langle \dot{x}^{\mu}(\tau_1) x^{\nu}(\tau_2)
\rangle \!\!&=&\!\!-g^{\mu\nu} \dot{G}_B(\tau_1,\tau_2) = -g^{\mu\nu}
\Bigl[ {\rm sign}(\tau_1-\tau_2)-{2(\tau_1-\tau_2)\over T}
\Bigr] \;,\nonumber\\
\langle \dot{x}^{\mu}(\tau_1) \dot{x}^{\nu}(\tau_2) \rangle \!\!&=&\!\!
+g^{\mu\nu} \ddot{G}_B(\tau_1,\tau_2) =
+g^{\mu\nu} \Bigl[2\delta(\tau_1-\tau_2)-{2\over T}\Bigr] \;.
\end{eqnarray}
(From these elementary prescriptions it is possible to derive
contraction rules for exponentials, which can be used equivalently.)
\par
{\it (ii) Integrations\/}: Next, the polynomial $\tau$-
and $\eta$-integrations have to be performed.
After the expansion in step {\it (i)} the integration over
the $\eta$'s is trivial.
The integrands of the
$\tau$-integrations are constructed from the worldline Green function
$G_B$ and its derivatives and are thus polynomials in the
variables $\tau_j$.
\par
{\it (iii) Cyclic reduction\/}: The number of terms will be
drastically reduced if one identifies structures which differ
only by cyclic permutation under the trace.
Since all cyclic permutations of a given term actually
occur during the calculation and since the polynomial integrations
in step {\it (ii)\/}
yield the same factor for terms equivalent under cyclic permutation
(due to the translational invariance of our Green function),
the identification of equivalent terms can easily be done.
\newpage
For the ungauged case, this already completes the program. Let us
therefore cite the results for a pure scalar background [7]. The
``master equation'' (\ref{master}) simplifies to
\begin{eqnarray}
\nonumber  \Gamma[V]
= \!\!\int_0^\infty {{\rm d}T \over T} \!\!\!\!\!
&[&\!\!\!\!\!\!4\pi T]^{-d/2}
{\rm e}^{-m^2T} {\rm tr} \int {\rm d}^dx_0 {\sum_{n=0}^\infty}
(-T)^n \int_0^{1} \!\!\!{\rm d}u_1\int_0^{u_1}
\!\!\!{\rm d}u_2 \,\;... \int_0^{u_{n-1}} \!\!\!\!\!{\rm d}u_n\\
&\times& {\rm exp} \Bigl[ - T {\sum_{i<k}} G(u_i,u_k)\del_{(i)}
\del_{(k)}\Bigr] V^{(1)}(x_0) ... V^{(n)}(x_0) \;,
\end{eqnarray}
where a rescaling to the unit circle, $\tau_i = T u_i\;(i=1,...,n)$,
has been performed using the scaling property of the Green function,
$G(\tau_1,\tau_2) = T G(u_1,u_2)$.The factor $[4\pi T]^{-d/2}$
arises from the normalization of the free path integral.
After expanding the exponential, performing the integrations and identifying
terms equivalent under cyclic permutation one obtains the final result in
the form
\begin{equation}
\Gamma[V] =
\int_0^\infty {{\rm d}T\over T} [4\pi T]^{-d/2} {\rm e}^{-m^2T}
{\sum_{n=1}^\infty} {(-T)^n\over n!} \!\!\int {\rm d}x_0
\;{\rm tr}\; O_n\;.
\end{equation}
The method was completely computerized using the algebraic
language FORM and -- for identification of cyclic redundancies
-- PERL. The calculation was performed up to $O_{11}$. In the following
the results to ${\cal O}(T^6)$ are quoted [8], using the shorthand notation
$V_{\kappa\lambda}\equiv\del_\kappa\del_\lambda V(x_0)$:
\begin{eqnarray*}
 O_1 \!\!\!\!\!&=&\!\!\!\!\!\! V ,\;\;
 O_2 \,= V^2 ,\;\;
 O_3 \,= \biggl( V^3 + {1\over 2} \Vka \Vka \biggr) ,\;\;
 O_4 \,= \biggl( V^4
 + 2 V \Vka \Vka + {1\over5} \Vkala \Vkala \biggr) ,
\cr\noalign{\vskip4pt}
 O_5 \!\!\!\!\!&=&\!\!\!\!\! \biggl( V^5 + 3 V^2 \Vka\Vka
 + 2 V\Vka V\Vka + V\Vkala\Vkala
 + {5\over 3}\Vka\Vla\Vkala + {1\over 14}\Vkalamu\Vkalamu\biggr)  \;\;,
\cr\noalign{\vskip4pt}
 O_6 \!\!\!\!\!&=&\!\!\!\!\!
 \biggl( V^6 + 4 V^3 \Vka \Vka + 6 V^2 \Vka V \Vka
 + {12\over7} V^2 \Vkala \Vkala + {9\over7} V \Vkala V \Vkala \cr
 + \!\!\!\!&&\!\!\!\!\!\!\!\!\!\!\! {26\over 7}V \Vkala\Vka\Vla
 + {26\over7} V \Vka \Vla \Vkala + {17\over14} \Vka\Vla\Vka\Vla
 + {18\over7} V \Vka \Vkala \Vla + {9\over7} \Vka\Vka\Vla\Vla \cr
 + \!\!\!\!&&\!\!\!\!\!\!\!\!\!\!\!{3\over7} V \Vkalamu \Vkalamu
 + \Vmu \Vkala \Vkalamu
 + \Vmu \Vkalamu \Vkala + {11\over21} \Vkala \Vlamu \Vkamu
 + {1\over42} \Vkalamunu \Vkalamunu \biggr)\;.
\end{eqnarray*}
For the higher order coefficients we simply quote the number of terms
in Table 1.
\par
\begin{table}[htbp]
\tcaption{Number of terms to $O(T^n)$ for ungauged case.}
\centerline{\vphantom{0}}
\centerline{\footnotesize\smalllineskip
\begin{tabular}{|@{\hspace{3mm}}c@{\hspace{3mm}}| c c c c c|}
\hline
\vphantom{\rule[-1.5mm]{0mm}{5mm}}n &7 &8 &9 &10 &11\\
\hline
\vphantom{\rule[-1.5mm]{0mm}{5mm}}\# of terms &37 &114 &380 &1373 &5301\\
\hline
\end{tabular}}
\end{table}
\par
It should be emphasized that no partial integrations have to be used
to obtain the result in a unique minimal basis, which does not contain
any box operators $\Box=\del^2$ (this is due to the fact that the Green
function obeys the property $G(u_i,u_i)=0$, i.e. there are no
self--contractions and box operators never occur).
\newpage
In a gauged theory, however, the worldline path integral method does
not yield a minimal basis after identification of cyclic permutations only.
This is due to Bianchi identities and the antisymmetry of the field
strength tensor. This gives additional relations among the
coefficients which have to be used to reduce the result
into a minimal basis. To show the
principle of the reduction scheme let us restrict to the pure gauge case
in the following. For a discussion of the general case including a proof
of minimality see [9]. Like in the pure scalar case there are no
self--contractions, i.e. the coefficients have the generic form
\begin{equation}
(D\ldots DF) \ldots (D_{\mu_1} \ldots D_{\mu_k} F_{\mu_{k+1}\mu_{k+2}})
\ldots (D\ldots DF) \;,
\end{equation}
where there is no contraction among the indices $\mu_i$ within a factor
$(D\ldots DF)$. Starting from this particular form of the coefficients the
reduction is performed with the following steps:
\par
{\it Step 1\/}: Consider terms with the following index structure:
\[
(D\ldots DF) \ldots (D_{\mu_1}\ldots D_{\mu_i} \ldots D_{\mu_k}
F_{\alpha\beta}) \ldots X_\alpha \ldots X_{\mu_i} \ldots X_{\beta} \ldots \;,
\]
where $X$ denotes $D$'s or $F$'s in different factors.
In a first step the derivatives are exchanged (producing additional terms
with a higher number of $F$'s) until $D_{\mu_i}$ acts on $F_{\alpha\beta}$
directly. Then the Bianchi identity and the antisymmetry of $F$ is used to
convert the term into two terms of the structure
\begin{eqnarray}
&&(D\ldots DF) \ldots (D_{\mu_1}\ldots D_{\mu_{i-1}} D_{\mu_{i+1}} \ldots
D_{\mu_k}D_\alpha F_{\mu_i\beta}) \ldots X_\alpha \ldots X_{\mu_i}
\ldots X_{\beta} \ldots \nonumber \\
&&(D\ldots DF) \ldots (D_{\mu_1}\ldots D_{\mu_{i-1}} D_{\mu_{i+1}} \ldots
D_{\mu_k}D_\beta F_{\alpha\mu_i}) \ldots X_\alpha \ldots X_{\mu_i}
\ldots X_{\beta} \ldots \,. \nonumber
\end{eqnarray}
In the first term the derivative $D$ acting on $F$ directly is contracted
with the first $X$, whereas it is contracted with the last $X$ in the
second term. This has to be done iteratively with all indices $\mu_i$ whose
contraction is ``nested'' between the contractions of $F$.
\par
{\it Step 2\/}: Multiple contractions between factors can be converted into
a standard form by exchanging $D$'s (again producing terms with a higher
number of $F$'s) and using Bianchi identities and the antisymmetry of $F$,
\begin{eqnarray}
&&(D\ldots DF) \ldots (D\ldots F_{\alpha\beta})\ldots (\ldots D_\alpha\ldots
D_\beta\ldots F_{\kappa\lambda}) \ldots (D\ldots DF) \;\longrightarrow
\nonumber \\
&&(D\ldots DF) \ldots (D\ldots F_{\alpha\beta})\ldots (\ldots D_\alpha\ldots
F_{\beta\lambda}) \ldots (D\ldots DF) \;\longrightarrow \nonumber \\
&&(D\ldots DF) \ldots (D\ldots F_{\alpha\beta}) \ldots (D \ldots
F_{\alpha\beta}) \ldots (D\ldots DF) \;, \nonumber
\end{eqnarray}
yielding a ``double contraction'' of $F$'s. However, this kind of reduction
is not always possible. There is a special case resulting in two terms:
\begin{eqnarray}
&&(D\ldots DF)\ldots (D\ldots D_\alpha F_{\beta\kappa})\ldots
(D\ldots D_\beta F_{\alpha\lambda}) \ldots (D\ldots DF)
\;\longrightarrow\nonumber \\
&&(D\ldots DF)\ldots (D\ldots D_\kappa F_{\alpha\beta})\ldots
(D\ldots D_\lambda F_{\alpha\beta}) \ldots (D\ldots DF) \;+\nonumber \\
&&(D\ldots DF)\ldots (D\ldots D_\alpha F_{\beta\kappa})\ldots
(D\ldots D_\alpha F_{\beta\lambda}) \ldots (D\ldots DF) \;. \nonumber
\end{eqnarray}
Again there is a double contraction in the first term. The contractions are
fixed such that the $D$'s and the first indices of the $F$'s are contracted
in the second term.
\par
{\it Step 3\/}: One can use the antisymmetry of $F$ and a
symmetrization/antisymmetri\-zation of the derivatives for a further
reduction. Consider a (remaining) index structure
\[
(D\ldots DF)\ldots(D\ldots D_\mu D_\nu F)\ldots
(D\ldots DF)(D\ldots DF_{\mu\nu})\ldots (D\ldots DF) \;.
\]
The antisymmetric part of the term results in a term of higher order in $F$,
whereas the symmetric part vanishes due to the antisymmetry of $F_{\mu\nu}$.
\par
{\it Step 4\/}: The last step is to fix the ordering of indices. For the
derivatives this can be done by exchanging them (producing terms with a higher
number of $F$'s again) whereas one can use the antisymmetry in case of the
$F$'s.
\par
Because terms with a smaller number of derivatives and higher number of field
strength tensors are produced throughout this procedure, one has to start the
reduction with the terms containing a maximum number of derivatives. To give
an example we quote the seven basis terms left in ${\cal O}(T^4)$ for the pure
gauge case:
\begin{eqnarray}
&&F_{\mu\nu}F_{\mu\nu}F_{\kappa\lambda}F_{\kappa\lambda} \;,\;
F_{\mu\nu}F_{\kappa\lambda}F_{\mu\nu}F_{\kappa\lambda} \;,\;
F_{\mu\nu}F_{\nu\kappa}F_{\kappa\lambda}F_{\lambda\mu} \;,\;
F_{\mu\kappa}F_{\kappa\nu}F_{\mu\lambda}F_{\lambda\nu} \;,\nonumber\\
&&D_\kappa F_{\mu\nu} D_\lambda F_{\mu\nu} F_{\kappa\lambda} \;,\;
D_\kappa F_{\mu\nu} D_\kappa F_{\mu\lambda} F_{\nu\lambda} \;,\;
D_\kappa D_\lambda F_{\mu\nu} D_\kappa D_\lambda F_{\mu\nu} \;.\nonumber
\end{eqnarray}
\par
In conclusion, we have obtained the inverse mass expansion of the one-loop
effective action within a worldline path integral formalism.
With a complete computerization of the algorithm we computed
the expansion coefficients up to order ${\cal O}(T^{11})$
for the ungauged case.
The application of the reduction scheme to the
coefficients in gauge theories is
in progress, as well as inclusion of background gravitational fields.
The method can be generalized to multiloop calculations in
scalar theories and quantum electrodynamics [10].
\par
\nonumsection{References}

\end{document}